\documentclass[baaa]{baaa}

\usepackage[pdftex]{hyperref}
\usepackage{subfigure}
\usepackage{natbib}
\usepackage{helvet,soul}
\usepackage[font=small]{caption}

\contriblanguage{1}

\contribtype{2}

\thematicarea{5}

\title{An analysis of the isomers HCN and HNC in the evolution of high-mass star-forming regions}

\titlerunning{HCN and HNC in high-mass star-forming regions}

\author{
N.C. Martinez\inst{1}, 
\&
S. Paron\inst{1}
}

\authorrunning{Martinez N. C. \& Paron S.}

\contact{nmartinez@iafe.uba.ar}

\institute{
Instituto de Astronomía y Física del Espacio, CONICET--UBA, Argentina
} 

\resumen{
El estudio de las moléculas y su química en regiones de formación estelar es crucial para entender los procesos físicos que allí ocurren. Las emisiones de la línea J=1--0 de los isómeros HCN y HNC fueron utilizadas para derivar sus intensidades integradas (I), probar una relación recientemente aparecida en la literatura entre temperatura cinética (T$_{K}$) y el cociente isomérico de las (I), y obtener abundancias (X) de estos isómeros en 55 regiones de formación estelar de alta masa. Estas últimas son clasificadas, de acuerdo a una escala evolutiva, en nubes oscuras infrarrojas, objetos protoestelares masivos, núcleos moleculares calientes, y regiones 
H\textsc{ii}~ultracompactas. Se infiere que las T$_{K}$ deducidas a partir del cociente de la intensidad integrada (I$^{HCN/HNC}$) son subestimadas, y por lo tanto sugerimos que dicha relación no puede ser empleada como un termómetro universal en el medio interestelar. Las abundancias de los isómeros muestran un comportamiento que se explica a través de la química que ocurre con el aumento de la temperatura y cantidad de radiación UV de acuerdo a las fases evolutivas. Encontramos que el cociente de abundancias (X$^{HCN/HNC}$) difícilmente puede ser utilizado como un reloj químico, y se sugiere que puede ser aproximado por el cociente I$^{HCN/HNC}$. 
Este trabajo es parte de un estudio en curso sobre múltiples moléculas que se encuentran en la muestra de regiones analizadas y pretende contribuir al conocimiento químico de la formación de estrellas de alta masa.
}

\abstract{
The study of molecules and their chemistry in star-forming regions is fundamental to understand the physical process occurring in such regions. 
The HCN and HNC J=1-0 emissions were used to derive their integrated line intensities (I), to probe a relation recently appeared in the literature between the kinetic temperatures (T$_{K}$) and the isomeric (I) ratio, and to obtain the isomers abundances (X) in 55 high-mass star-forming regions. These last ones are classified, according to the evolutive stage, as infrared dark clouds, high-mass protostellar objects, hot molecular cores, and ultracompact H\textsc{ii}~regions. It is inferred that the T$_{K}$ obtained from the isomeric integrated intensity ratio 
(I$^{HCN/HNC}$) are underestimated, and hence we suggest that this relation cannot be employed as an universal thermometer in the interstellar medium. The isomers abundances show a behavior that can be explained from the chemistry occurring as the temperature and the UV radiation increase according to the evolutive stage. We found that the abundance ratio (X$^{HCN/HNC}$) hardly could be used as a chemical clock, and we suggest that it can be approximated by I$^{HCN/HNC}$. 
This work is part of an on-going study of multiple molecules that stand in the sample of analyzed regions which intends to contribute in the chemical knowledge of high-mass star formation.
}

\keywords{
Stars: formation --- ISM: molecules
}

\begin{document}

\maketitle

\section{Introduction}

Molecular gas plays a  major role in star formation and the emission of many molecules are commonly used to examine the environment around star-forming regions. Given that such molecules are ubiquitous in all the phases that a young massive stellar object (MYSO) goes through (subjected to different conditions of pressure, temperature, and density), the chemistry that arises is very rich and diverse, making its study of vital significance for understanding the birth and evolution of high-mass stars.
In particular, the constitutional isomers hydrogen cyanide and isocyanide (HCN and HNC, respectively) are two of the simplest molecules found in the interstellar medium (ISM). Although they have very similar physical parameters (close emission frequencies, similar rotational constants, Einstein coefficients, etc.) and an interconnected chemistry, usually differences in their spatial distribution within a molecular cloud may reflect the chemical conditions of the gas and the evolution of the star-forming regions \citep{schilke92}.

Recently, \citet{hcn} proposed the integrated intensity ratio of the isomers as an useful thermometer to estimate the kinetic temperature (T$_{\rm K}$) of the molecular gas. The authors studied the emission of such molecules throughout the Integral Shape Filament (ISF) in Orion, and developed an empirical correlation between the HCN/HNC ratio and T$_{\rm K}$. They suggested that this new tool can be used in other regions of the ISM. With the aim of investigating the validity of such thermometer in high-mass star-forming regions and studying other parameters that can be derived from the HCN and HNC emissions, we present a spectroscopic study, from a chemical perspective, of 55 sources in different stages of star formation.

\section{Data and analyzed sources}

The analyzed HCN J=1--0 molecular line contains three hyperfine components: F=1--1 at 88.6304 GHz, F=2--1 at 88.6318 GHz and F=0--1 at 88.6339 GHz; the HNC J=1--0 line has a rest frequency at 90.6635 GHz. The data were obtained from the catalogue J/A+A/563/A97 in the ViZieR database\footnote{http://cdsarc.u-strasbg.fr/viz-bin/qcat?J/A+A/563/A97}, which were obtained and studied by \citet{gerner}.
Following the sample of sources included in the above mentioned catalogue, we investigate the molecular gas conditions and chemistry of HCN and HNC related to infrared dark clouds (IRDC), high-mass protostellar objects (HMPO), hot molecular cores (HMC), and ultracompact H\textsc{ii}~regions (UCH\textsc{ii}). 
The number of sources in each category is 19, 20, 7, and 9 respectively. The data, obtained with the 30\,m IRAM telescope, are described in detail in \citet{gerner}. 

\section{Results}

\subsection{HCN/HNC ratio and T$_{\rm K}$}\label{hacar}

\begin{figure*}
    \centering
    \includegraphics[width=5.3cm]{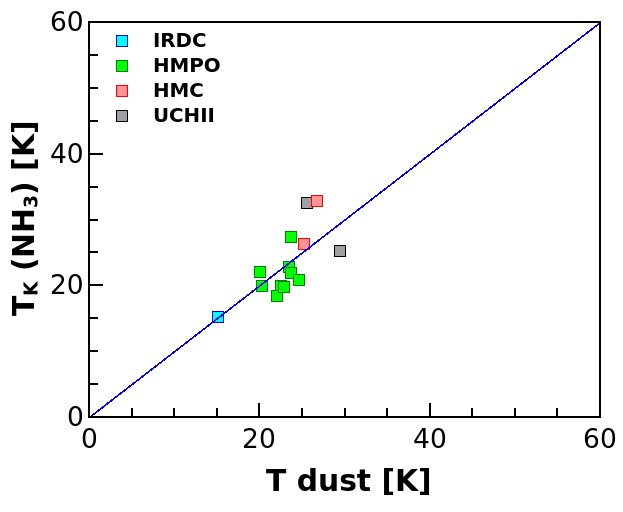}
    \includegraphics[width=5.3cm]{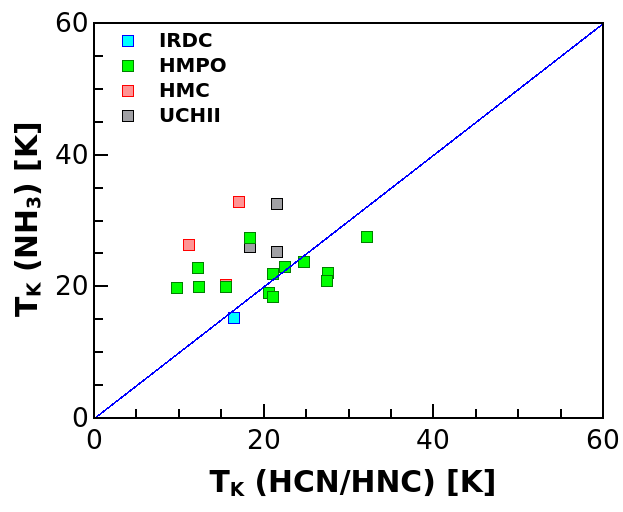}
    \includegraphics[width=5.3cm]{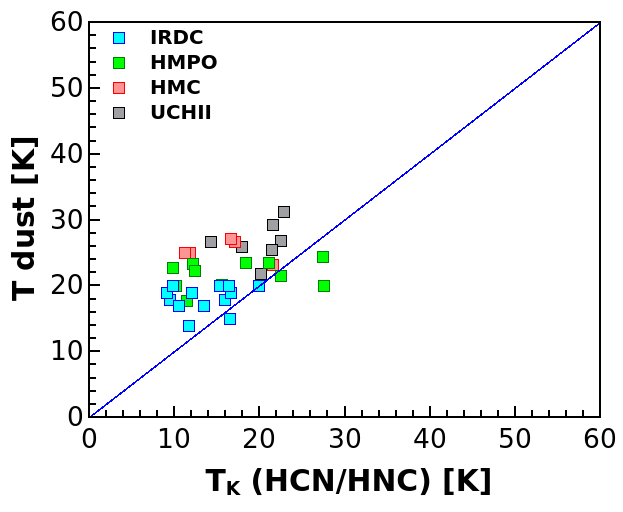}
    \caption{\emph{Left:} Kinetic temperature of ammonia vs. dust temperature obtained from the PPMAP procedure done to the Hi-GAL maps \citep{marsh}. \emph{Center:} Kinetic temperature of ammonia vs. kinetic temperature derived from the HCN-HNC integrated intensity ratio. \emph{Right:} Dust temperature vs. kinetic temperature derived from the HCN-HNC integrated intensity ratio. The blue line in all cases indicates the unity.}
    \label{comparacion}
\end{figure*}

We calculated the T$_{\rm K}$ for each source using the correlation presented by \citet{hcn}.
As done by the authors, we integrated the HCN and HNC J=1--0 line including all hyperfine components. For reasons of space, we only display the average kinetic temperature for each kind of source in Table\,\ref{deltaT} (Col.\,2). With the aim of testing the robustness of this isomeric thermometer, we compared the derived values of T$_{\rm K}$(HCN/HNC) with dust temperature ($\rm {T_{dust}}$) and ammonia kinetic temperature ($\rm {T_{K}(NH_{3})}$) in sources that these data were available. $\rm {T_{dust}}$ values were obtained from the maps\footnote{http://www.astro.cardiff.ac.uk/research/ViaLactea/} generated by the PPMAP procedure done to the Hi-GAL maps in the wavelength range 70–500 $\mu$m \citep{marsh}. $\rm {T_{K}(NH_{3})}$ was extracted from \cite{urqu11} (see equations 3, 4 and references therein), who calculate this parameter in several sources here included and which are part of the Red MSX Source survey \citep{urqu08}. We show average $\rm {T_{dust}}$ and $\rm {T_{K}(NH_{3})}$ values for each kind of source in Table \ref{deltaT} (Cols.\,3 and 4, respectively). Figure \ref{comparacion} exhibits the comparison between T$_{\rm K}$(HCN/HNC) and $\rm {T_{K}(NH_{3})}$ (central panel) and $\rm {T_{dust}}$ (right panel). In Fig.\,\ref{comparacion}, left panel, the correlation between $\rm {T_{dust}}$ and $\rm {T_{K}(NH_{3})}$ can be appreciated. Typical errors in HCN and HNC integrated intensities are about 0.2 and 0.1 K km s$^{-1}$ respectively, which yields errors between 1\% and 5\% in the T$_{\rm {K}}$(HCN/HNC).

\subsection{HCN and HNC abundances}\label{3.2}

Additionally, to investigate the chemistry of HCN and HNC along the presented sources representing different stages of star-forming evolution, we calculated the abundances of both molecules in each source. In the absence of the isotopic species regarding these two species, we estimated the total column densities of these isomers (${N_{tot}}$), according to Equation \ref{nu} (\citealp{tielens, mangum}):
\begin{align}
    \label{nu}
N_{tot}& =\frac{1.94\times10^{3}}{A_{ul}} \frac{Q~{\nu^{2}_{ul}}}{g_{u}~exp[-E_{u}/kT_{ex}]} \int T_{mb}~ dv
\end{align}
\noindent where $A_{ul}$ is the Einstein coefficient in s$^{-1}$, $\nu_{ul}$ is the line frequency in GHz, the integrated intensity is in K km s$^{-1}$, Q is the partition function, $g_{u}$ and $E_{u}$ are the weight and the state energy both of the upper level, respectively, and $T_{ex}$~is the excitation temperature, which is assumed to be equal to $T_{K}$, and we adopted $T_{ex}$ = $T_{dust}$ by assuming that gas and dust are thermally coupled (see Section\,\ref{thermo}). The use of these formulae assumes 
local thermodynamic equilibrium (LTE), that both isomers have the same $T_{ex}$~and are optically thin. This last one can be in general a strong assumption. Opacities effects will be analyzed in a forthcoming work. Here it is used as a rough estimation.
 
To obtain the isomers abundances (X$^{\rm HCN}$, X$^{\rm HNC}$) their column densities were divided by the H$_{2}$ column density extracted from \citet{marsh} in each source. The average abundances are shown in Table\,\ref{abundances}. To interpret these parameters according to the evolutionary trend, we performed the abundance ratio between hydrogen cyanide and isocyanide (X$^{\rm HCN/HNC}$) which was evaluated along with the ratio of the integrated intensities (I$^{\rm HCN/HNC}$, previously calculated in the Sect.\,\ref{hacar}). These results are presented in Table\,\ref{comparison}.

\begin{table}
\centering
\caption{Calculated average temperatures.}
\label{deltaT}
\begin{tabular}{lccc}
\hline\hline
\noalign{\smallskip}
Source	&	$\overline{\rm T_{K} (HCN/HNC)}$	&	$\overline{\rm T_{dust}}$	&	$\overline{\rm T_{K}(NH_{3})}$	\\
	&	(K)	&	(K)	&	(K)	\\
\hline
IRDC	&	14.9$\pm$1.3	&	18.5$\pm$0.4	&	15.3$^{*}$	\\
HMPO	&	21.9$\pm$1.9	&	21.8$\pm$0.6	&	22.2$\pm$0.8	\\
HMC	&	15.6$\pm$1.6	&	26.0$\pm$0.6	&	26.6$\pm$3.7	\\
UCH\textsc{ii}	&	19.8$\pm$1.0	&	26.3$\pm$1.1	&	28.0$\pm$2.3	\\

\hline
\multicolumn{4}{l}{$^{*}$Value obtained from a single source.}
\end{tabular}
\end{table}

\section{Discussion}

\subsection{Employing an isomeric relation to achieve kinetic temperatures}
\label{thermo}

From the the comparison between ${\rm T_{K}(NH_{3})}$ and ${\rm T_{dust}}$ (see Fig.\,\ref{comparacion}, left panel), even though not all sources have information on dust temperature and ammonia ${\rm T_{K}}$ at the same time, it can be appreciated a very good correlation between these temperatures, indicating that molecular gas and dust are indeed thermally coupled. We suggest that this condition can be extrapolated to the whole sample of sources supporting the validity of the comparison between ${\rm T_{K}(HCN/HNC)}$ and the ${\rm T_{dust}}$.

Figure\,\ref{comparacion} (central and right panels) show an underestimation in ${\rm T_{K}}$ derived from the isomers, which is even stronger when compared with ${\rm T_{dust}}$. In particular, in the average values obtained from each kind of source (see Table\,\ref{deltaT}), this is more notorious towards HMCs and UCH\textsc{ii}~ regions. This issue could be due to opacity effects in the HCN line, which it is being studied and the results will be presented in a forthcoming work.  
Therefore, we suggest that the empirical thermometer based on the isomeric ratio could be quite reliable on IRDC and HMPO stages, and we cannot conclude that  ${\rm I^{HCN/HNC}}$ can be used efficiently as a good proxy for the ${\rm T_{K}}$ in general in the ISM. Certainly, more analysis and studies to discern this topic are needed.

It is important to notice that the average ${\rm T_{dust}}$ and ${\rm T_{K}(NH_{3})}$ values increase with the star-forming region evolutive scenario: they go from $\sim$18 K in IRDC, to $\sim$22 K in HMPO, reaching finally an approximate value of 26 K in HMC and UCH\textsc{ii}~stages. This is consistent with an increment of the temperature that occurs continuously with the star forming processes and this is observed in the behavior of ${\rm T_{K} (HCN/HNC)}$ at the very first two stages.

\begin{table}
\centering
\caption{Calculated isomer average abundances. Values displayed as $a(x)=a\times10^{x}$.}
\label{abundances}
\begin{tabular}{lcc}
\hline\hline
Source	&	$\overline{\rm X^{HCN}}$	&	$\overline{\rm X^{HNC}}$	\\
\hline
IRDC	&	5.6(-10)$\pm$1.1(-10)	&	4.2(-10)$\pm$6.1(-11)	\\
HMPO	&	2.2(-09)$\pm$4.2(-10)	&	1.2(-09)$\pm$2.9(-10)	\\
HMC	&	5.9(-09)$\pm$3.0(-10)	&	3.6(-09)$\pm$1.5(-09)	\\
UCH\textsc{ii}	&	3.0(-09)$\pm$6.4(-10)	&	1.6(-09)$\pm$2.9(-10)	\\

\hline
\end{tabular}
\end{table}

\begin{table}
\centering

\caption{Average abundance and integrated intensity ratios between the isomers.}
\label{comparison}
\begin{tabular}{lcc}
\hline\hline
Source	&	$\overline{\rm X^{HCN/HNC}}$	&	$\overline{\rm I^{HCN/HNC}}$	\\
\hline
IRDC	&	1.4$\pm$0.5	&	1.3$\pm$0.2	\\
HMPO	&	1.8$\pm$0.8	&	2.1$\pm$0.2	\\
HMC	&	1.6$\pm$0.7	&	1.4$\pm$0.2	\\
UCH\textsc{ii}	&	1.9$\pm$0.8	&	1.8$\pm$0.2	\\

\hline
\end{tabular}
\end{table}

\subsection{The behavior of HCN and HNC through the star-forming evolution}
 
Regarding the abundances of these two species for each star-forming phase, it can be seen in Table\,\ref{abundances} that there is an increment towards the HMC phase of both molecules, and then fall slightly towards the UCH\textsc{ii}~state. We infer that this behavior may be due to the progressive formation of the isomers from IRDC to the HMC stage where the temperature allows the chemistry to proceed and be rich. Then, the destruction of both species occurs due to much higher temperature and the surrounding radiation when the stars born.
When evaluating the ratio of abundances presented in Table\,\ref{comparison}, it is noted that it rises marginally as temperature rises in agreement with what was presented by \cite{goldsmith}. In fact, ${\rm X^{HCN/HNC}}$ follows the same trend presented in the ratio of the integrated intensities. We suggest that this global behavior is a consequence of the favored destruction of HNC at high temperatures, where transfers to HCN, causing an increasing ratio value \citep{graninger}. However, the results for the ratio of abundances overlap and are not significant, which do not allow us to use this parameter as a chemical clock.
 
Our results agree with those presented by \cite{hoq} (measurement of ${\rm I^{HCN/HNC}}$) but are contrary to those presented by \cite{jin} who do find ${\rm X^{HCN/HNC}}$ as an indicator of the evolutionary status. A possible explanation for this might be that the authors used the isotopic lines H$^{13}$CN and HN$^{13}$C and perform a wide analysis regarding the optical depth.
Indeed, more analysis on a new number of line-rich sources on different formation stages is necessary to study this behavior and resolve this issue. Finally, according to our calculations, we also propose that ${\rm I^{HCN/HNC}}$ can roughly be used to estimate the abundance ratio (see Table\,\ref{comparison}), which is actually useful because such an estimate is based on a direct measurement instead on a dedicated calculus that uses some assumptions. 

Studying the emission and the chemistry of HCN and HNC is very useful to derive physical parameters that are necessary to study high-mass star formation. Studies like presented here are important due to the statistical information that can be obtained. Increasing the samples of the analyzed sources is planed.

\begin{acknowledgement}
N.C.M. is a doctoral fellow of CONICET, Argentina.
S.P. is member of the {\sl Carrera del Investigador Cient\'\i fico} of CONICET, Argentina.
This work was partially supported by the Argentina grant PIP 2021 11220200100012 from CONICET.
\end{acknowledgement}


\bibliographystyle{baaa}
\small
\bibliography{ref}
 
\end{document}